\documentclass[10pt,conference]{IEEEtran}
\usepackage{graphicx,
	psfrag,
	epsfig,
	epstopdf,
    enumerate,
	amsthm,
    amsmath,
	amssymb,
	url,
	algorithm,
	algorithmic,
	balance,
	enumerate,
	color,
    url,
	setspace,
	tikz,
	pgfplots,
    geometry,
    placeins,
    caption,
    cleveref,
    multirow,
    array,
}
\usepackage{slashbox}
\usepackage{amsmath}
\usepackage{cite}
\newgeometry{left=1.5cm,right=1.5cm,bottom=1.5cm, top=1.5cm}

\newcommand{\ignore}[1]{}

\begin{document}
\IEEEoverridecommandlockouts
\IEEEpubid{\makebox[\columnwidth]{978-1-5386-4889-6/18/\$31.00 \copyright 2018 IEEE \hfill} \hspace{\columnsep}\makebox[\columnwidth]{ }}
\title{Relationship between Student Engagement and Performance in e-Learning Environment Using Association Rules}	
\author{
\IEEEauthorblockN{Abdallah Moubayed, MohammadNoor Injadat, Abdallah Shami, and Hanan Lutfiyya}
\IEEEauthorblockA{ Western University, London, Ontario, Canada \\
	e-mails: \{amoubaye, minjadat, Abdallah.Shami, hlutfiyy\}@uwo.ca,
}	
}
\maketitle
\begin{abstract}
The field of e-learning has emerged as a topic of interest in academia due to the increased ease of accessing the Internet using using smart-phones and wireless devices. One of the challenges facing e-learning platforms is how to keep students motivated and engaged. Moreover, it is also crucial to identify the students that might need help in order to make sure their academic performance doesn't suffer. To that end, this paper tries to investigate the relationship between student engagement and their academic performance. Apriori association rules algorithm is used to derive a set of rules that relate student engagement to academic performance. Experimental results' analysis done using confidence and lift metrics show that a positive correlation exists between students' engagement level and their academic performance in a blended e-learning environment. In particular, it is shown that higher engagement often leads to better academic performance. This cements the previous work that linked engagement and academic performance in traditional classrooms.
\end{abstract}
\begin{IEEEkeywords}
	e-Learning, Association Rules, Apriori Algorithm, Support, Confidence, Lift
\end{IEEEkeywords}
\vspace{-3mm}
\section{Introduction}\label{intro}
\indent With the proliferation of technology and the expanding use of smart devices, the field of e-learning has grown in popularity in recent times. E-learning can be defined to be the use of electronic devices and technology for learning new information and skills \cite{e-learning}. Such environments allow individuals to learn new skills without having a physical mentor teaching them face-to-face. In layman's terms, e-learning can be defined to be the use of electronic technologies to access to educational curriculum outside of a traditional classroom \cite{e-learning}. Many e-learning platforms have been proposed such as Coursera, edX, and OWL in which courses are offered fully online. Moreover, several prestigious universities such as Stanford University and Massachusetts Institute of technology have also started offering online courses through their own platforms known as Stanford Online \cite{Stanford_online} and MIT OpenCourseWare respectively \cite{MIT_online}.\\
\indent One of the challenges an e-learning environment faces is how to keep students motivated and not allow them to become isolated. This is particularly of importance in online courses since statistics have shown that on average only around \textbf{15-25\%} of students complete online courses they registered for \cite{mooc_dropout_rate1}\cite{mooc_dropout_rate2}. Therefore, keeping students motivated is crucial. This is because students might feel discouraged if they perceive that their learning pace is slower than others, especially with little to no face-to-face interaction with instructors or fellow students. Moreover, studies have shown that engagement with course material has an important impact on future career decisions taken by students \cite{e-learning_motivation2}. Therefore, teachers need to find a way to keep students engaged and motivated \cite{engagement1,engagement2,engagement3}. Furthermore, many literature works have shown a positive relation between student engagement and academic performance with higher engagement level associated with better grades \cite{engagement_and_performance1} \cite{engagement_and_performance2}.\\
\indent In this paper, the relationship between the students' engagement level and their academic performance is studied. Despite there being a few literature works that study this relationship \cite{engagement_and_performance3} \cite{engagement_and_performance4}, none of the previous works consider a comprehensive set of engagement metrics and their impact on academic performance. This paper studies the impact of both the frequency-related and time-related engagement metrics as well as the overall engagement level on the academic performance of students using Apriori association rules algorithm.\\
\indent The remainder of this paper is organized as follows: Section \ref{background} gives a brief background about the field of e-Learning and the association rules. Then, Section \ref{related} presents some of the related work from the literature. Section \ref{system_model} gives overview of the system model considered in this work. That is followed by Section \ref{data_desc} which describes the dataset available and its transformation into the considered features. Section \ref{performance} discusses the experiments conducted and the resulting rules obtained based on two evaluation metrics. Finally, Section \ref{conc} concludes the paper.
\section{Background}\label{background}
\subsection{E-learning:}
\indent E-learning has become more popular in recent times due to the proliferation of technology throughout the world and the boom in information access. This is because it allows individuals to learn new skills without having to engage in a physical face-to-face teaching environment. In layman's terms, e-learning can be defined to be the access to educational curriculum outside of a traditional classroom by utilizing electronic technologies \cite{e-learning}. Several more complex definitions have been given for e-learning \cite{e-learning_def}. However, they all agree on one common point which is the use of technology and technological devices such as computers and handheld devices as a means to access and share information \cite{e-learning_def}. \\
\indent This can shown how the field of e-learning can significantly contribute to the notion of big data. With increasing number of students accessing educational material online, more data flows are generated. Thus, machine learning and data analytics also become crucial in order to make use of the growing amount of collected data generated in the field of e-learning. In particular, the sub-field of educational data mining has emerged which focused on analyzing educational data to better understand and improve students' performance as well as enhance the learning and teaching process \cite{EDM1} \cite{EDM2}.
\subsection{Association Rules:}
\indent Association rule learning is a type of rule-based machine learning algorithms that aims to discover interesting relations between items in large databases \cite{association_def1}. The idea is to produce rules that can predict the occurrence of an item based on occurrences of other items \cite{association_def2}. Agrawal \textit{et al.} provide a more formal definition as follows \cite{association_def3}:\\
Let $I=\{i_1,i_2,...,i_n\}$ be a set of $n$ attributes/items and $T=\{t_1,t_2,...,t_m\}$ be a set of $m$ transactions forming a database. Each transaction $t_i$ includes a subset of the items available in $I$. A rule can be defined as $X \implies Y$ where $X,Y \subseteq I$, i.e. $X$ and $Y$ (known as itemsets) are a subsets of the items available. In other words, a rule can be thought of as a predictable transaction within the database. $X$ and $Y$ are commonly referred to as the antecedent and the consequent of the rule respectively.\\
\indent Association rules can be beneficial in an e-learning environment as they can detect correlations between different features within the dataset. In particular, they can be used to correlate student behaviors with their performance to determine what is positively or negatively impacting their learning experience.\\
\indent In order to evaluate the importance and interestingness of an association rule, several measures have been proposed. In what follows, three measures are presented.
\begin{itemize}
	\item \textbf{Support:} Support of an itemset is an indication of how frequent an itemset appears in the transactions' database. It can be thought of as the probability of occurrence of the considered itemset by counting the number of transactions in which the itemset appears relative to the total number of transactions. More formally, the support of an itemset $X$ with respect to a set of transactions $T$ is calculated as \cite{association_def2}:
	\begin{equation}
	supp(X)= \frac{|{t \in T; X \subseteq t}|}{|T|}
	\end{equation}  
	\item \textbf{Confidence:} Confidence is a measure of how frequent the rule is within the transactions' database. In layman terms, it is the portion of transactions that contain both itemsets $X$ and $Y$ forming the rule relative to the transactions that contain $X$ in general. Hence, the confidence of rule $X \implies Y$ can be defined as \cite{association_def2}:
	 \begin{equation}
	 confidence(X \implies Y)= \frac{supp(X \cup Y)}{supp(X)}
	 \end{equation}
	 The confidence of a rule can be thought of as the conditional probability of the rule \cite{association_eval1}.
	\item \textbf{Lift:} The lift of a rule is a measure of how interesting the rule is. The lift determines the probability of the rule occurring relative to the probability of the antecedent and consequent being independent. Therefore, the lift of an association rule is defined as \cite{association_eval2}:
	\begin{equation}
	lift(X \implies Y)= \frac{supp(X \cup Y)}{supp(X) \times supp(Y)}
	\end{equation}
	A lift of 1 implies that the two itemsets comprising the rule are independent and hence the rule associating them together is not truly a rule. However, if the lift is $>$ 1, it can be concluded that the two itemsets are dependent on each other. This makes the corresponding rule possibly useful in predicting future occurrences of the consequent if the antecedent occurs.\\
	The significance of the lift is that it takes into consideration both the confidence of the rule as well as the overall transaction database \cite{association_eval3}.
\end{itemize}
\indent \indent Association rules are often used for categorical (non-numeric) datasets and have been considered for a variety of applications. This includes marketing and sales promotion, supermarket shelf management, inventory management \cite{association_def2}, network intrusion detection \cite{association_app1}, and health informatics \cite{association_app2}. \\
\indent Several association rules algorithms have been developed. In what follows, a brief description of three well known algorithms is given.   
\subsubsection{Apriori Algorithm}\mbox{}\\
\indent Apriori algorithm is a breadth-first search-based algorithm that depends on the frequency of the itemsets to identify a set of association rules \cite{Apriori1}. The algorithm identifies the itemsets that appear in at least $C$ transactions within the database, where $C$ is the minimum threshold chosen by the user. It adopts a "bottom up" approach where a frequent subset is extended by one item in each iteration. This means that the algorithm starts with itemsets of length 1 (i.e. only one item within the itemset) and determines the itemset that have a frequency higher than the considered threshold $C$. This is repeated until no more new frequent itemsets are found. The length is then incremented by 1 and the same process is adopted again. This continues until there is no more possible extensions of the itemsets.\\
\indent The popularity of the apriori algorithm stems from the fact that it can be easily implemented and parallelized as well as makes use of the large itemset property \cite{Apriori_adv_disadv}. However, the algorithm does suffer from some drawbacks. One of the main drawbacks is that it requires several database scans in order to produce the rules \cite{Apriori_adv_disadv}. This is mainly due to the itemset extension property of the algorithm. Another drawback is the fact that it can be slow due to its dependence on the size of the database, number of items within it, and the choice of minimum support \cite{association_def2}. 
\subsubsection{Frequent Pattern (FP)-Growth Algorithm}\mbox{}\\
\indent Frequent pattern growth algorithm, also known as FP-growth algorithm, is an efficient and scalable alternative to the apriori algorithm \cite{FPgrowth1}. The FP-growth algorithm also aims to find the frequent itemsets within the transaction database. However, it does so without using candidate generations, i.e. it doesn't increment the itemset size by 1 in each round. This helps it outperform the apriori algorithm in terms of runtime as it significantly reduces the number of considered frequent itemsets. In essence, this algorithm adopts a divide-and-conquer approach  to generate the association rules \cite{FPgrowth1}. The algorithm is based on the use of a particular data structure named the FP-tree which conserves the itemset association information \cite{FPgrowth1}. To produce the FP-tree, the transaction database is scanned once and the set of frequent items $F$ is then determined and ordered in support-descending order. Then each transaction within the database is considered with the items in it sorted based on the order found in $F$. The count of each item in the tree is updated with each new transaction studied. Hence, the FP-tree is produced using two scans of the database \cite{FPgrowth1}.\\
\indent Despite its efficiency and runtime speed, this algorithm does suffer from one major drawback. That drawback is that all attributes of the database should be binomial, i.e. either it exists in the transaction or not \cite{FPgrowth_disadv}. Hence, it can't be used in cases where an attribute/item can have multiple possible values.
\subsubsection{Generalized Sequential Pattern Algorithm (GSP)}\mbox{}\\
\indent Generalized Sequential Pattern (GSP) algorithm is another association rule algorithm. It is mainly used for sequential mining. That means it looks for patterns in datasets in which values are given in a sequential manner \cite{GSP1}. The algorithm uses the apriori algorithm level-wise to discover the frequent itemsets within the corresponding level. The set of frequent itemset is then given to the GSP algorithm which makes multiple scans in order to determine the association rules. In each scan, a set of candidate sequences of length $k$ are formed using itemsets of length $(k-1)$. This continues until no more frequent sequences can be generated.\\
\indent The GSP algorithm suffers from similar limitations as its apriori counterpart. One major limitation is that it needs multiple database scans to generate the association rules \cite{GSP_disadv}. This is especially a concern when dealing within large transaction databases. Another limitation is the generation of non-existent candidates. This happens because the GSP algorithm generates candidate sequences by combining smaller ones without accessing the database. This can lead to time wasted when considering non-existent patterns \cite{GSP_disadv}.
\section{Related Work \& Contribution}\label{related}
\subsection{Student Engagement:}
\indent Several works in the literature have explored the notion of student engagement levels and methodologies on how to define and determine these levels \cite{engagement_level1}\cite{engagement_level2}. Oriogun classified students based on their engagement into three levels, namely High, Nominal, and Low \cite{engagement_level1}. The authors used the SQUAD approach in which a student's posts on a discussion forum were classified as one of five categories, namely: Suggestion, Question, Unclassified, Answer, and Delivery. Each post belonging to one of the aforementioned categories was given a score and the total score of each student was used to determine their engagement in the course. However, this study only focused on quality of forum posts to evaluate engagement without taking into consideration other possible metrics. Kamath \textit{et al.} also classified students using a three-level model. However, the authors used image recognition as the basis of their classification by building a custom dataset of images portraying the different engagement levels and using support vector machines to classify new images \cite{engagement_level2}. Yet, this provides a limitation as not all students will have cameras in their devices. Moreover, this won't be as beneficial in a non-real time scenario.\\
\indent To be able to measure/identify the engagement level of students, different engagement metrics have been used and proposed in the literature \cite{engagement_metrics1,engagement_metrics3,engagement_metrics4}. Reid used the "Classroom Survey of Student Engagement (CLASSE)" model and proposed frequency-related metrics to determine the engagement level of the students \cite{engagement_metrics1}. The metrics considered were: number of questions asked, number of participations, number of interactions with instructor, and time spent on the course.
On the other hand, Ramesh \textit{et al.} used three main frequency-related features to determine the engagement level of students in a Massive Open Online Course (MOOC) setting \cite{engagement_metrics4}. The engagement metrics are number of posts in discussion forums, number of content views, and binary indicator of assignment completion. These metrics were used to determine the students' engagement based on a three-level model. However, most previous works lack the use of a comprehensive set of engagement metrics that include both frequency-related and time-related metrics.
\vspace{-3mm}
\subsection{Academic Performance Prediciton:}
\indent Similarly, several literature works tried to predict the academic performance of students based on a variety of factors. Yadav \textit{et al.} studied the impact of previous or first year exams marks on the final grade of engineering students \cite{performance1}. In their experiment, the authors used several classification algorithms and found that the C4.5 (also known as J48) algorithm gives the most accurate results. This was corroborated by Ramesh \textit{et al.}'s work in \cite{performance2} which also concluded that decision tree classification algorithm was the most suitable for predicting students’ performance. Moreover, it was shown that students’ previous data can be used to predict their final grade. Prasad \textit{et al.} also considered the use of decision tree algorithms and found that the C4.5 (J48) algorithm is the most suitable algorithm to predict students’ grades because of its accuracy and speed \cite{performance3}. 
\subsection{Impact of Engagement on Academic Performance:}
\indent Some literature works have also studied the impact of student engagement on their academic performance \cite{engagement_and_performance1} \cite{engagement_and_performance2} \cite{engagement_and_performance3}. Lee studied the relationship between student engagement and their academic performance using the data from 121 U.S. schools \cite{engagement_and_performance1}. The study concluded that both behavioral and emotional engagement can be used to predict reading performance accurately. On the other hand, Casuso-Holgado \textit{et al.} used a questionnaire that aims to measure the students engagement and investigated the relationship with academic achievement. The study concluded that engagement is positively related with the academic achievement of students. Similarly, Madar \textit{et al.} aimed to study the relationship between the student engagement through the use of an e-learning platform and their academic performance. This was done using a survey questionnaire filled by postgraduate students. Results showed a positive correlation between engagement on the platform and the performance of students. \\
\indent However, most works that studied this relationship didn't consider it for an e-learning environment. Moreover, the few works that did investigate this relationship in an e-learning setting didn't use a comprehensive set of engagement metrics to fully understand their impact on the academic performance of students.
\subsection{Contribution:}
\indent The contribution of this work can be summarized as follows:
\begin{itemize}
	\item Providing a comprehensive set of engagement metrics that include both frequency-related metrics and time spent on different tasks of the course.
	\item Studying the impact of the considered engagement metrics and levels on academic performance.
\end{itemize}
\section{System Model}\label{system_model}
\begin{figure}[htbp!]
	\centering
	\includegraphics[scale=0.5]{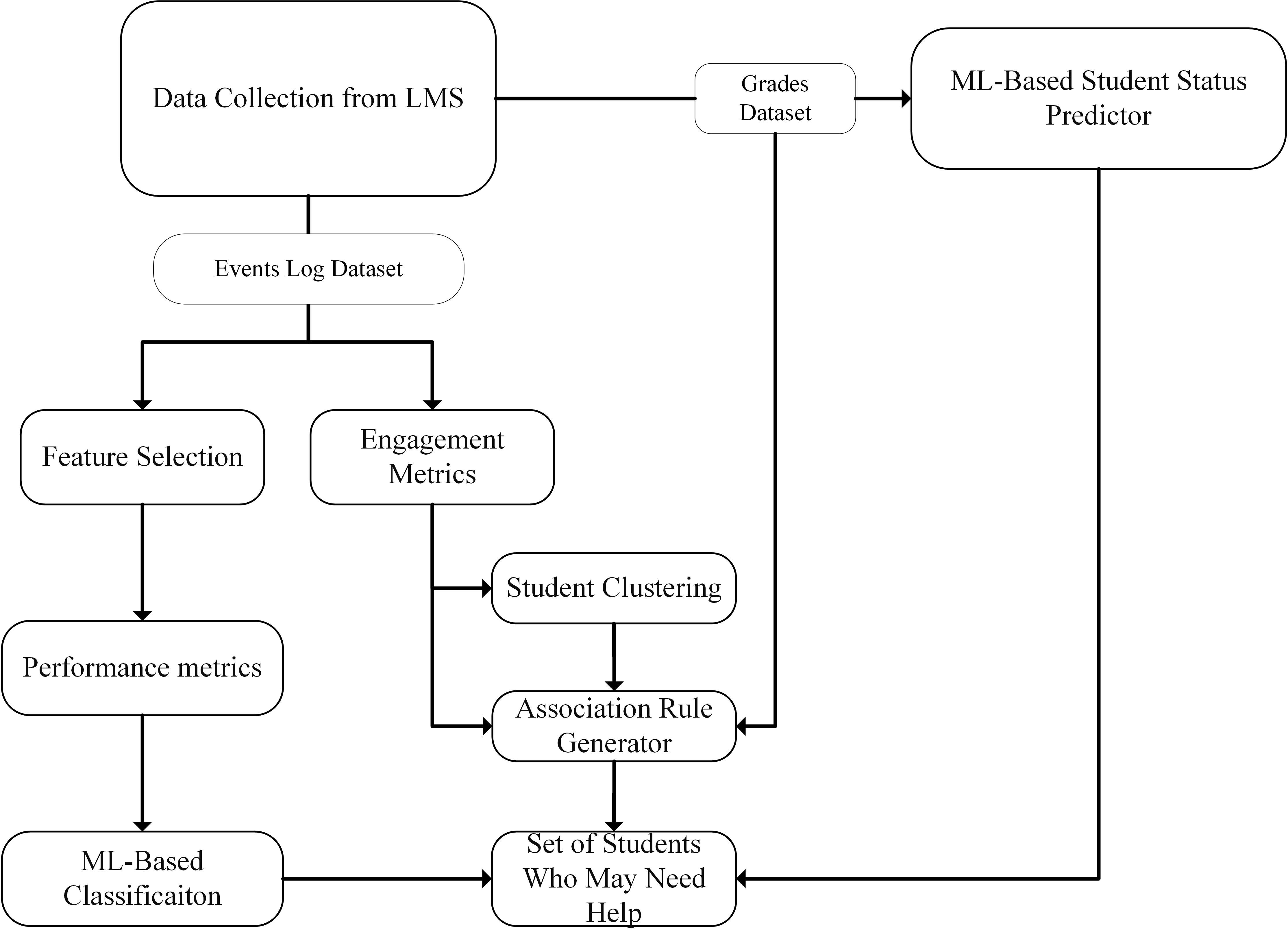}
	\caption{LMS Analytical Module}\label{analytical_module}
\end{figure}

\indent Figure \ref{analytical_module} presents the overall system model considered in this work. The presented module is the general module that would be integrated into the LMS. It consists of the following modules:
\begin{itemize}
	\item Data Collection Module: Collects the data from the LMS. This includes both the events and the grades.
	\item ML-based Student Status Predictor Module: Predicts the status of the students using different supervised learning algorithms.
	\item Engagement Metrics Module: Calculates the different engagement metrics using the collected events log data.
	\item Student Clustering Module: Clusters students into different engagement levels using the aforementioned engagement metrics by implementing unsupervised learning algorithms.
	\item Association Rule Generator Module: Takes the both the grades dataset and calculated engagement metrics as well as the resulting clustering decision to generate a group of association rules.
	\item Feature Selection Module: Selects the features that are most representative of the student performance.
	\item Performance Metrics Module: Generates a group of performance metrics that can be used to classify students. 
	\item ML-based Classification Module: Classifies students by implementing different supervised learning algorithms.
	\item Set of Students Who May Need Help Module: Identifies the group of students who may need help in the course. This can be used as an initial step of the e-learning personalization process.
\end{itemize}
\indent\indent This work focuses on the association rule generator module. Within this module, several different association rule algorithms can be implemented. In this case, the apriori algorithm is used. This is because as discussed previously, the FP-growth algorithm only accepts binomial features. On the other hand, the GSP algorithm searches for association rules in a transactional dataset. However, the dataset considered here is not a transactional one, i.e. each record represents one student. Hence, the GSP algorithm can't be implemented.
\section{Dataset Description}\label{data_desc}
\subsection{Data Preprocessing:}
\indent The collected dataset is from a second year undergraduate Science course offered in a North American University. The dataset consists of two parts. The first part is an event log of 486 enrolled students and has a total of 305933 records collected from the university's learning management system (LMS). Each record has the following 6 fields:
\begin{itemize}
	\item Event Date: The time-stamp of the event.
	\item Event Type: The type of action the student makes.
	\item Event Location: The directory in which the action was taken by the student.
	\item Session Start: The time-stamp signaling the start of the online session.
	\item Session End: The time-stamp signaling the end of the online session.
	\item Student ID: Student Identifier.
\end{itemize}
The dataset was sorted based on the ``Student ID'' first and then based on ``Event Date''. This was done so that a chronological order of the events each student took is obtained.\\
\indent The second part is the obtained grades of the 486 students in the different assignments, quizzes, and exams. Each record has the following 8 fields:
\begin{itemize}
	\item Student ID: Student Identifier.
	\item Assingment 1: Grade obtained in Assignment 1.
	\item Assingment 2: Grade obtained in Assignment 2.
	\item Assingment 3: Grade obtained in Assignment 3.
	\item Quiz 1: Grade obtained in Quiz 1.
	\item Midterm Exam: Grade obtained in Midterm Exam.
	\item Final Exam: Grade obtained in Final Exam.
	\item Course Grade: Final Course Grade.
\end{itemize} 
\subsection{Data Transformation:}
\indent Using MATLAB, the event log dataset was transformed from its original state into a new dataset representing the students' engagement metrics. This was done by searching the events' column of the subset of data representing each student and calculating the engagement metrics accordingly. Moreover, each metric's value was rounded to the nearest 10s to get a discrete set of values as per the requirement of the different association rules algorithms. Furthermore, the engagement level of each student is determined as per our previous work \cite{Abdallah_kmeans}. Based on the engagement metrics used in the literature and the original available dataset, Table \ref{table_of_metrics} presents the calculated engagement metrics and engagement level which are considered in this study as well as their description, type, and range of values of each metric. Note that for the engagement level, L means Low while M means Medium and H means High.\\
\begin{table}[ht!]
	\centering
	\caption{Engagement Metrics Description}
	\scalebox{0.75}{
		\begin{tabular}{|p{1.6cm}|l|p{1.2cm}|p{2cm}|}
			\hline
			Feature & Description & Type & Range of Values \\ \hline
			Student ID&Student identifier&Numeric&[0,...,485]\\ \hline
			Num. of Logins& \begin{minipage}[t]{0.57\columnwidth}The number of times the student accessed the course site on the LMS\end{minipage}&Numeric&[0,10,...,650]\\ \hline
			Num. of Content Reads&\begin{minipage}[t]{0.55\columnwidth} The number of times the student accessed/downloaded course material\end{minipage}&Numeric&[0,10,...,1010]\\ \hline
			Num. of Forum Reads&\begin{minipage}[t]{0.55\columnwidth} The number of times the student read posts on the discussion forum\end{minipage}&Numeric&[0,10,...,60]\\ \hline
			Num. of Forum Posts&\begin{minipage}[t]{0.55\columnwidth} The number of times the student posted on the discussion forum\end{minipage}&Numeric&[0,10]\\ \hline
			Num. of Quiz Reviews&\begin{minipage}[t]{0.55\columnwidth} The number of times the student reviewed their quiz solution before final submission\end{minipage}&Numeric&[0,10]\\ \hline
			Assign. 1 duration to submit (in hours)&\begin{minipage}[t]{0.55\columnwidth} The duration (in hours) between Assignment 1 posting and submission\end{minipage}&Numeric& [0,10,...,580]\\ \hline
			Assign. 2 duration to submit (in hours)&\begin{minipage}[t]{0.55\columnwidth} The duration (in hours) between Assignment 2 posting and submission\end{minipage}&Numeric& [0,10,...,300]\\ \hline
			Assign. 3 duration to submit (in hours)&\begin{minipage}[t]{0.55\columnwidth} The duration (in hours) between Assignment 3 posting and submission\end{minipage}&Numeric& [0,10,...,630]\\ \hline
			Average Assign. duration to submit (in hours)&\begin{minipage}[t]{0.55\columnwidth} The average duration (in hours) between Assignments' posting and submission\end{minipage}&Numeric& [0,10,...,500]\\ \hline
			Engagement Level&\begin{minipage}[t]{0.55\columnwidth} Student engagement level using K-means Clustering\end{minipage}&Categoric& [L,M,H]\\ \hline
	\end{tabular}}
	\label{table_of_metrics}
\end{table}
\indent Similarly, the grades of each student were rounded to the nearest 10s to get a discrete set of values as per the requirement of the association rules algorithms. The calculated engagement metrics dataset and the grades dataset were combined to form a new dataset consisting of eighteen features, namely the Student ID, nine engagement metrics, engagement level, and the seven grades.
\section{Performance Evaluation \& Discussion}\label{performance}
\subsection{Experiment Setup}\label{experiment_setup}
\indent As mentioned earlier, the experiment uses apriori algorithm to generate the association rules linking the students' engagement metrics and level to their performance. To do so, the following settings were used:
\subsubsection{Software}\mbox{}\\
\indent The following software have been used to run the experiment and record the results:
\begin{itemize}
	\item Operating System: Microsoft Windows 10 (64-Bit OS, X-64 based processor)
	\item MATLAB: MATLAB was used to transform the data from its original state to the new desired dataset that includes the engagement metrics as as the grades.
	\item Waikato Environment for Knowledge Analysis (WEKA) version 3.8: WEKA is a toolkit used for machine learning and data mining. It was developed using Java by the University Waikato in New Zealand.
\end{itemize}
\subsubsection{Experiment}\mbox{}\\
\indent WEKA was used to run the apriori algorithm to generate the desired association rules. Apriori algorithm was used because it allows features to have multiple values rather than just binomial as is the case with FP-growth algorithm. Furthermore, apriori algorithm was used instead of GSP algorithm because the considered dataset is not transactional in nature, i.e. each student has only one record within the dataset. Hence, GSP algorithm can't be implemented. 
\subsection{Rules:}\label{results}
\indent To generate the association rules, the minimum support used was 0.1, i.e. the rule needs to have occurred at least 10\% of the time to be considered. Moreover, the minimum confidence value assumed was 0.9, i.e. we would want to be more than 90\% confident that the rule is applicable. These values were chosen to ensure that the rules are frequent and interesting enough to be considered in an educational setting \cite{min_conf}. Based on the aforementioned settings, the following rules have been generated:  
\begin{itemize}
	\item \textbf{Engagement level=H \& Quiz1$\geq$90$\implies$Course Grade$\geq$ 90:}\\
	The rule had a support = 0.2, i.e. 20\% of all students were classified as highly engaged, had a grade higher than 90 in their quiz, and completed the course with a grade higher than 90. The rule had a confidence of 1 and a lift of 4.02. This means that all the students who were highly engaged and performed well in their quiz ended up excelling in the course. The high lift value shows that the components are highly correlated. This means that engagement level can indeed be used as a predictor for academic performance.
	\item \textbf{Number of Logins$\geq$60 \& Quiz1$\geq$80$\implies$ 
		Course Grade $\geq$ 80:}\\
	The support of this rule is 0.61, which indicates that around 61\% of the students logged into the system more than 60 times. This is an indication of being moderately or highly engaged as per the results obtained in \cite{Abdallah_kmeans}. The rule's confidence is 1 and the lift is 1.37, indicating that all students who were moderately or highly engaged and performed well in their quiz had a high final course grade. Moreover, the lift value indicates that there is a positive correlation between the antecedent and consequent of the rule.
	\item \textbf{Engagement level=M \&  Quiz1$\geq$70$\implies$ 70$\leq$Course Grade$\leq$90:}\\
	This rule had a support of 0.4, confidence of 0.93, and a lift of 1.1. This shows again that there is a correlation between engagement and academic performance. This is because students that were moderately engaged tended to get a moderate grade in the course. 
	\item \textbf{Engagement level=M $\implies$ 70$\leq$Course Grade$\leq$90:}\\
	confidence =0.9 lift = 1.1
	This rule further emphasizes the previous obtained rule. The rule had a support of 0.4 as well. The confidence is 0.9 and the lift is 1.1 too. 
\end{itemize}
\indent In addition to the obtained rules, the average grade of the students is calculated based on their engagement. It can be shown that students with medium engagement had an average final grade of 77 while those that were highly engaged had an average final grade of 79. On the other hand, students that were lowly engaged had an average grade of 61. This cements the hypothesis that student engagement is highly correlated with academic performance even in a blended e-learning environment. Hence, it can be used as a predictor and identifier of students that may need help based on their engagement in the course.
\section{Conclusion}\label{conc}
\indent The field of e-learning has become more popular due to the proliferation of technology. Among the challenges facing e-learning is keeping the students engaged and motivated. This is especially important due to the impact that engagement has on academic performance. To that end, this paper proposed the use of apriori algorithm to investigate the impact of engagement on students' academic performance in a blended e-learning environment. Nine engagement metrics that include both frequency-related as well as time-related metrics were identified and calculated from the event log dataset available and combined with the grades to form a new dataset composed of eighteen features. Apriori algorithm was used to generate rules that studied the impact of students' engagement level on their academic performance. Experimental results showed that there is a high positive correlation between engagement and academic performance. This was shown through the association rules generated which stated with high confidence and lift that students with higher levels of engagement tended to perform better in the course. Due to this positive correlation, engagement can be used as a predictor of the students' academic performance. This in turn can be used to identify the unengaged students who may need help with the course and work with them to improve their engagement and possibly their performance.

\small
\bibliographystyle{IEEEtran}
\bibliography{Ref}

\begin{thebibliography}{10}
\providecommand{\url}[1]{#1}
\csname url@samestyle\endcsname
\providecommand{\newblock}{\relax}
\providecommand{\bibinfo}[2]{#2}
\providecommand{\BIBentrySTDinterwordspacing}{\spaceskip=0pt\relax}
\providecommand{\BIBentryALTinterwordstretchfactor}{4}
\providecommand{\BIBentryALTinterwordspacing}{\spaceskip=\fontdimen2\font plus
\BIBentryALTinterwordstretchfactor\fontdimen3\font minus
  \fontdimen4\font\relax}
\providecommand{\BIBforeignlanguage}[2]{{%
\expandafter\ifx\csname l@#1\endcsname\relax
\typeout{** WARNING: IEEEtran.bst: No hyphenation pattern has been}%
\typeout{** loaded for the language `#1'. Using the pattern for}%
\typeout{** the default language instead.}%
\else
\language=\csname l@#1\endcsname
\fi
#2}}
\providecommand{\BIBdecl}{\relax}
\BIBdecl

\bibitem{e-learning}
E.~Inc., ``{About E-Learning},'' Available at:
  \url{http://www.elearningnc.gov/about_elearning/}, 2016.

\bibitem{Stanford_online}
{Stanford University}, ``{Stanford Online},'' Available at:
  \url{https://online.stanford.edu}.

\bibitem{MIT_online}
{Massachusetts Institute of technology}, ``{MIT OpenCourseWare},'' Available
  at: \url{https://ocw.mit.edu/index.htm}.

\bibitem{mooc_dropout_rate1}
K.~Jordan, ``{MOOC Completion Rates:The Data},'' Available at:
  \url{http://www.katyjordan.com/MOOCproject.html}, 2015.

\bibitem{mooc_dropout_rate2}
F.~Manjoo, ``{Udacity Says It Can Teach Tech Skills to Millions, and Fast},''
  Available at:
  \url{https://www.nytimes.com/2015/09/17/technology/udacity-says-it-can-teach-tech-skills-to-millions.html?_r=0},
  2015.

\bibitem{e-learning_motivation2}
K.~Kori, M.~Pedaste, H.~Altin, E.~Tõnisson, and T.~Palts, ``{Factors That
  Influence Students Motivation to Start and to Continue Studying Information
  Technology in Estonia},'' \emph{IEEE Transactions on Education}, vol.~59,
  no.~4, pp. 255--262, Nov 2016.

\bibitem{engagement1}
M.~Highley, ``{e-Learning: Challenges and Solutions},'' Available at:
  \url{https://elearningindustry.com/e-learning-challenges-and-solutions},
  2014.

\bibitem{engagement2}
M.~Handelsman, W.~Briggs, N.~Sullivan, and A.~Towler, ``{A measure of college
  student course engagement},'' \emph{The Journal of Educational Research},
  vol.~98, no.~3, pp. 184--192, 2005.

\bibitem{engagement3}
M.-T. Wang and J.~S. Eccles, ``{School context, achievement motivation, and
  academic engagement: A longitudinal study of school engagement using a
  multidimensional perspective},'' \emph{The Journal of Learning and
  Instruction}, vol.~28, pp. 12--23, 2013.

\bibitem{engagement_and_performance1}
\BIBentryALTinterwordspacing
J.-S. Lee, ``The relationship between student engagement and academic
  performance: Is it a myth or reality?'' \emph{The Journal of Educational
  Research}, vol. 107, no.~3, pp. 177--185, 2014. [Online]. Available:
  \url{http://dx.doi.org/10.1080/00220671.2013.807491}
\BIBentrySTDinterwordspacing

\bibitem{engagement_and_performance2}
\BIBentryALTinterwordspacing
M.~J. Casuso-Holgado, A.~I. Cuesta-Vargas, N.~Moreno-Morales, M.~T.
  Labajos-Manzanares, F.~J. Bar{\'o}n-L{\'o}pez, and M.~Vega-Cuesta, ``The
  association between academic engagement and achievement in health sciences
  students,'' \emph{BMC Medical Education}, vol.~13, no.~1, p.~33, 2013.
  [Online]. Available: \url{http://dx.doi.org/10.1186/1472-6920-13-33}
\BIBentrySTDinterwordspacing

\bibitem{engagement_and_performance3}
M.~J. Madar and O.~B. Ibrahim, ``E-learning towards student academic
  performance,'' in \emph{2011 International Conference on Research and
  Innovation in Information Systems}, Nov. 2011, pp. 1--5.

\bibitem{engagement_and_performance4}
T.~Rodgers, ``{Student Engagement in e-Learning Process and the Impact on Their
  Grades},'' \emph{International Journal of Cyber Society and Education},
  vol.~1, no.~2, pp. 143--156, Mar. 2008.

\bibitem{e-learning_def}
\BIBentryALTinterwordspacing
A.~Sangrà, D.~Vlachopoulos, and N.~Cabrera, ``{Building an Inclusive
  Definition of E-learning: An Approach to The Conceptual Framework},''
  \emph{The International Review of Research in Open and Distributed Learning},
  vol.~13, no.~2, pp. 145--159, 2012. [Online]. Available:
  \url{http://www.irrodl.org/index.php/irrodl/article/view/1161}
\BIBentrySTDinterwordspacing

\bibitem{EDM1}
G.~Kaur and W.~Singh, ``{Prediction Of Student Performance Using Weka Tool},''
  \emph{{An International Journal of Engineering Sciences}}, vol.~17, pp.
  8--16, Jan. 2016.

\bibitem{EDM2}
A.~Azis, N.~Ismail, and F.~Ahmad, ``{Mining Students' Academic Performance},''
  \emph{{Journal of Theoretical and Applied Information Technology}}, 2013.

\bibitem{association_def1}
G.~Piatetsky-Shapiro, ``Discovery, analysis and presentation of strong rules,''
  in \emph{Knowledge Discovery in Databases}, G.~Piatetsky-Shapiro and W.~J.
  Frawley, Eds.\hskip 1em plus 0.5em minus 0.4em\relax AAAI Press, 1991, pp.
  229--248.

\bibitem{association_def2}
{L. Antonie}, ``{INTRODUCTION TO DATA MINING ASSOCIATION RULES},'' {{University
  of Guelph}}.

\bibitem{association_def3}
\BIBentryALTinterwordspacing
R.~Agrawal, T.~Imieli\'{n}ski, and A.~Swami, ``Mining association rules between
  sets of items in large databases,'' in \emph{Proceedings of the 1993 ACM
  SIGMOD International Conference on Management of Data}, ser. SIGMOD
  '93.\hskip 1em plus 0.5em minus 0.4em\relax New York, NY, USA: ACM, 1993, pp.
  207--216. [Online]. Available: \url{http://doi.acm.org/10.1145/170035.170072}
\BIBentrySTDinterwordspacing

\bibitem{association_eval1}
J.~Hipp, U.~G\"{u}ntzer, and G.~Nakhaeizadeh, ``Algorithms for association rule
  mining \&mdash; a general survey and comparison,'' \emph{SIGKDD Explor.
  Newsl.}

\bibitem{association_eval2}
S.~Brin, R.~Motwani, J.~D. Ullman, and S.~Tsur, ``Dynamic itemset counting and
  implication rules for market basket data,'' in \emph{ACM SIGMOD International
  Conference on Management of Data (SIGMOD '97)}.

\bibitem{association_eval3}
\BIBentryALTinterwordspacing
M.~Hahsler, B.~Gruen, and K.~Hornik, ``Introduction to arules - {A}
  computational environment for mining association rules and frequent item
  sets,'' \emph{Journal of Statistical Software}, vol.~14, no.~15, pp. 1--25,
  Oct. 2005. [Online]. Available: \url{http://dx.doi.org/10.18637/jss.v014.i15}
\BIBentrySTDinterwordspacing

\bibitem{association_app1}
Y.~Changguo, W.~Nianzhong, W.~Tailei, Z.~Qin, and Z.~Xiaorong, ``The research
  on the application of association rules mining algorithm in network intrusion
  detection,'' in \emph{2009 First International Workshop on Education
  Technology and Computer Science}, vol.~2, Mar. 2009, pp. 849--852.

\bibitem{association_app2}
W.~Altaf, M.~Shahbaz, and A.~Guergachi, ``Applications of association rule
  mining in health informatics: a survey,'' \emph{Artificial Intelligence
  Review}, vol.~47, no.~3, pp. 313--340, Mar 2017.

\bibitem{Apriori1}
R.~Agrawal and R.~Srikant, ``Fast algorithms for mining association rules,'' in
  \emph{Proc. of 20th Intl. Conf. on VLDB}, 1994, pp. 487--499.

\bibitem{Apriori_adv_disadv}
I.~S. of~Engineering, ``Apriori algorithm,'' Available at:
  \url{https://www.slideshare.net/INSOFE/apriori-algorithm-36054672}, 2014.

\bibitem{FPgrowth1}
\BIBentryALTinterwordspacing
J.~Han, J.~Pei, and Y.~Yin, ``Mining frequent patterns without candidate
  generation,'' in \emph{Proceedings of the 2000 ACM SIGMOD International
  Conference on Management of Data}, ser. SIGMOD '00.\hskip 1em plus 0.5em
  minus 0.4em\relax New York, NY, USA: ACM, 2000, pp. 1--12. [Online].
  Available: \url{http://doi.acm.org/10.1145/342009.335372}
\BIBentrySTDinterwordspacing

\bibitem{FPgrowth_disadv}
Rapidminer, ``Fp-growth,'' Available at:
  \url{https://docs.rapidminer.com/studio/operators/modeling/associations/fp_growth.html}.

\bibitem{GSP1}
N.~R. Mabroukeh and C.~I. Ezeife, ``A taxonomy of sequential pattern mining
  algorithms,'' \emph{ACM Computing Surv.}

\bibitem{GSP_disadv}
P.~Fournier-Viger, J.~C.-W. Lin, R.~U. Kiran, and Y.~S. Koh, ``A survey of
  sequential pattern mining,'' \emph{Data Science and Pattern Recognition},
  vol.~1, no.~1, pp. 54--77, 2017.

\bibitem{engagement_level1}
P.~K. Oriogun, ``{Towards understanding online learning levels of engagement
  using the SQUAD approach to CMC Discourse},'' \emph{Australian Journal of
  Educational Technology}, vol.~19, pp. 371--387, Nov. 2003.

\bibitem{engagement_level2}
A.~Kamath, A.~Biswas, and V.~Balasubramanian, ``A crowdsourced approach to
  student engagement recognition in e-learning environments,'' in \emph{IEEE
  Winter Conference on Applications of Computer Vision (WACV'16)}, Mar. 2016,
  pp. 1--9.

\bibitem{engagement_metrics1}
L.~F. Reid, ``{Redesigning a Large Lecture Course for Student Engagement:
  Process and Outcomes},'' \emph{The Canadian Journal for the Scholarship of
  Teaching and Learning}, vol.~3.

\bibitem{engagement_metrics3}
A.~Koster, T.~Primo, A.~Oliveira, and F.~Koch, ``Toward measuring student
  engagement: a data-driven approach,'' in \emph{13th International Conference
  on Intelligent Tutoring Systems (ITS'16)}, Jun. 2016, pp. 60--68.

\bibitem{engagement_metrics4}
A.~Ramesh, D.~Goldwasser, B.~Huang, H.~Daum{\'e}, and L.~Getoor, ``Modeling
  learner engagement in moocs using probabilistic soft logic,'' in
  \emph{Advances in Neural Information Processing Systems Workshop on Data
  Driven Education}, 2013.

\bibitem{performance1}
\BIBentryALTinterwordspacing
S.~K. Yadav and S.~Pal, ``Data mining: {A} prediction for performance
  improvement of engineering students using classification,'' \emph{World of
  Computer Science and Information Technology Journal}, vol.~2, 2012. [Online].
  Available: \url{http://arxiv.org/abs/1203.3832}
\BIBentrySTDinterwordspacing

\bibitem{performance2}
V.~Ramesh, P.~Parkavi, and K.~Ramar, ``Predictive student performance: {A}
  statistical and data mining approach,'' \emph{International Journal of
  computer application}, vol.~63, no.~8, pp. 35--43, 2013.

\bibitem{performance3}
G.~Prasad and A.~Babu, ``Mining previous marks data to predict students'
  performance in their final year examination,'' \emph{International Journal of
  engineering research and technology}, vol.~2, no.~2, pp. 1--4, 2013.

\bibitem{Abdallah_kmeans}
A.~Moubayed, M.~Injadat, A.~Shami, and H.~Lutfiyya, ``{Student Engagement Level
  in e-learning Environment: Clustering Using K-means},'' \emph{{Elsevier
  Computers and Education (Under Review)}}.

\bibitem{min_conf}
\BIBentryALTinterwordspacing
S.~Ougiaroglou and G.~Paschalis, \emph{Association Rules Mining from the
  Educational Data of ESOG Web-Based Application}.\hskip 1em plus 0.5em minus
  0.4em\relax Berlin, Heidelberg: Springer Berlin Heidelberg, 2012, pp.
  105--114. [Online]. Available:
  \url{https://doi.org/10.1007/978-3-642-33412-2_11}
\BIBentrySTDinterwordspacing

\end{thebibliography}
\end{document}